\documentclass[conference]{IEEEtran}
\ifCLASSINFOpdf
\else
\fi
\usepackage{multirow}
\usepackage{subcaption}
\usepackage{graphicx}

\usepackage{fancyhdr}

\begin{document}
%
\title{Sentiment Analysis of Twitter Data for\\Predicting Stock Market Movements}

\author{\IEEEauthorblockN{Venkata Sasank Pagolu}
\IEEEauthorblockA{
School of Electrical Sciences\\
Computer Science and  Engineering\\
Indian Institute of Technology,\\
Bhubaneswar, India 751013 \\
Email: vp12@iitbbs.ac.in}
\and

\IEEEauthorblockN{Kamal Nayan Reddy Challa}
\IEEEauthorblockA{
School of Electrical Sciences\\
Computer Science and Engineering\\
Indian Institute of Technology,\\
Bhubaneswar, India 751013\\
Email: kc11@iitbbs.ac.in}
\and
\IEEEauthorblockN{Ganapati Panda}
\IEEEauthorblockA{
School of Electrical Sciences\\
Indian Institute of Technology\\
Bhubaneswar, India 751013\\
Email: gpanda@iitbbs.ac.in}
\and
\IEEEauthorblockN{\hspace{20pt}Babita Majhi}
\IEEEauthorblockA{
	\hspace{25pt}Department of Computer Science and IT\\
	\hspace{25pt}G.G Vishwavidyalaya, Central University \\
\hspace{25pt}Bilaspur, India 495009\\
\hspace{25pt}Email: babita.majhi@gmail.com}
}

\maketitle
\thispagestyle{fancy}


%


\maketitle

\begin{abstract}
Predicting stock market movements is a well-known problem of interest. Now-a-days social media is perfectly representing the public sentiment and opinion about current events. Especially, twitter has attracted a lot of attention from researchers for studying the public sentiments. Stock market prediction on the basis of public sentiments expressed on twitter has been an intriguing field of research. Previous studies have concluded that the aggregate public mood collected from twitter may well be correlated with Dow Jones Industrial Average Index (DJIA). The thesis of this work is to observe how well the changes in stock prices of a company, the rises and falls, are correlated with the public opinions being expressed in tweets about that company. Understanding author's opinion from a piece of text is the objective of sentiment analysis. The present paper have employed two different textual representations, Word2vec and N-gram, for analyzing the public sentiments in tweets. In this paper, we have applied sentiment analysis and supervised machine learning principles to the tweets extracted from twitter and analyze the correlation between stock market movements of a company and sentiments in tweets. In an elaborate way, positive news and tweets in social media about a company would definitely encourage people to invest in the stocks of that company and as a result the stock price of that company would increase. At the end of the paper, it is shown that a strong correlation exists between the rise and falls in stock prices with the public sentiments in tweets.
\end{abstract}


%
\IEEEpeerreviewmaketitle

\section{Introduction}
Earlier studies on stock market prediction are based on the historical stock prices. Later studies have debunked the approach of predicting stock market movements using historical prices. Stock market prices are largely fluctuating. The efficient market hypothesis (EMH) states that financial market movements depend on news, current events and product releases and all these factors will have a significant impact on a company's stock value~\cite{Fama}. Because of the lying unpredictability in news and current events, stock market prices follow a random walk pattern and cannot be predicted with more than 50\% accuracy~\cite{Qian}.

With the advent of social media, the information about public feelings has become abundant. Social media is transforming like a perfect platform to share public emotions about any topic and has a significant impact on overall public opinion. Twitter, a social media platform, has received a lot of attention from researchers in the recent times. Twitter is a micro-blogging application that allows users to follow and comment other users thoughts or share their opinions in real time~\cite{Adamic}. More than million users post over 140 million tweets every day. This situation makes Twitter like a corpus with valuable data for researchers~\cite{Jansen}.Each tweet is of 140 characters long and speaks public opinion on a topic concisely. The information exploited from tweets are very useful for making predictions~\cite{Pak}.

In this paper, we contribute to the field of sentiment analysis of twitter data. Sentiment classification is the task of judging opinion in a piece of text as positive, negative or neutral.

There are many studies involving twitter as a major source for public-opinion analysis. Asur and Huberman~\cite{Asur} have predicted box office collections for a movie prior to its release based on public sentiment related to movies, as expressed on Twitter. Google flu trends are being widely studied along with twitter for early prediction of disease outbreaks. Eiji et al.~\cite{Eiji} have studied the twitter data for catching the flu outbreaks. Ruiz et al.~\cite{Ruiz} have used time-constrained graphs to study the problem of correlating the Twitter micro-blogging activity with changes in stock prices and trading volumes. Bordino et al. ~\cite{Bordino} have shown that trading volumes of stocks traded in NASDAQ-100 are correlated with their query volumes (i.e., the number of users requests submitted to search engines on the Internet). Gilbert and Karahalios~\cite{Gilbert} have found out that increases in expressions of anxiety, worry and fear in weblogs predict downward pressure on the S\&P 500 index. Bollen~\cite{Bollen} showed that public mood analyzed through twitter feeds is well correlated with Dow Jones Industrial Average (DJIA). All these studies showcased twitter as a valuable source and a powerful tool for conducting studies and making predictions.

Rest of the paper is organized as follows. Section 2 describes the related works and Section 3 discusses the data portion demonstrating the data collection and pre-processing part. In Section 4 we discuss the sentiment analysis part in our work followed by Section 5 which examines the correlation part of extracted sentiment with stocks. In Section 6 we present the results, accuracy and precision of our sentiment analyzer followed by the accuracy of correlation analyzer. In Section 7 we present our conclusions and Section 8 deals with our future work plan.


\hfill 
 
\hfill 

\section{Related Work}

The most well-known publication in this area is by Bollen~\cite{Bollen}. They investigated whether the collective mood states of public (Happy, calm, Anxiety) derived from twitter feeds are correlated to the value of the Dow Jones Industrial Index. They used a Fuzzy neural network for their prediction. Their results show that public mood states in twitter are strongly correlated with Dow Jones Industrial Index. Chen and Lazer~\cite{Chen} derived investment strategies by observing and classifying the twitter feeds. Bing et al.~\cite{Bing} studied the tweets and concluded the predictability of stock prices based on the type of industry like Finance, IT etc. Zhang~\cite{Zhang} found out a high negative correlation between mood states like hope, fear and worry in tweets with the Dow Jones Average Index. Recently, Brian et al.~\cite{Brian} investigated the correlation of sentiments of public with stock increase and decreases using Pearson correlation coefficient for stocks. In this paper, we took a novel approach of predicting rise and fall in stock prices based on the sentiments extracted from twitter to find the correlation. The core contribution of our work is the development of a sentiment analyzer which works better than the one in Brian's work and a novel approach to find the correlation. Sentiment analyzer is used to classify the sentiments in tweets extracted.The human annotated dataset in our work is also exhaustive. We have shown that a strong correlation exists between twitter sentiments and the next day stock prices in the results section. We did so by considering the tweets and stock opening and closing prices of Microsoft over a year.

\section{Data Collection and Preprocessing}

\subsection{Data Collection}
A total of 2,50,000 tweets over a period of August 31st, 2015 to August 25th,2016 on Microsoft are extracted from twitter API~\cite{twitterapi}. Twitter4J is a java application which helps us to extract tweets from twitter. The tweets were collected using Twitter API and filtered using keywords like \$ MSFT, \# Microsoft, \#Windows etc. Not only the opinion of public about the company's stock but also the opinions about products and services offered by the company would have a significant impact and are worth studying. Based on this principle, the keywords used for filtering are devised with extensive care and tweets are extracted in such a way that they represent the exact emotions of public about Microsoft over a period of time. The news on twitter about Microsoft and tweets regarding the product releases were also included. Stock opening and closing prices of Microsoft from August 31st, 2015 to August 25th, 2016 are obtained from Yahoo! Finance~\cite{yahoo}.

\subsection{Data Pre-Processing}
Stock prices data collected is not complete understandably because of weekends and public holidays when the stock market does not function. The missing data is approximated using a simple technique by Goel~\cite{Goel}. Stock data usually follows a concave function. So, if the stock value on a day is x and the next value present is y with some missing in between. The first missing value is approximated to be (y+x)/2 and the same method is followed to fill all the gaps.

Tweets consists of many acronyms, emoticons and unnecessary data like pictures and URL's. So tweets are preprocessed to represent correct emotions of public. For preprocessing of tweets we employed three stages of filtering: Tokenization, Stopwords removal and regex matching for removing special characters.

\subsubsection{Tokenization}
Tweets are split into individual words based on the space and irrelevant symbols like emoticons are removed. We form a list of individual words for each tweet.

\subsubsection{Stopword Removal}
Words that do not express any emotion are called Stopwords. After splitting a tweet, words like a,is, the, with etc. are removed from the list of words.

\subsubsection{Regex Matching for special character Removal}
Regex matching in Python is performed to match URL’s and are replaced by the term URL. Often tweets consists of hashtags(\#) and @ addressing other users. They are also replaced suitably. For example, \#Microsoft is replaced with Microsoft and @Billgates is replaced with USER. Prolonged word showing intense emotions like coooooooool! is replaced with cool! After these stages the tweets are ready for  sentiment classification.

\begin{figure*}[h]
  \centering
  \includegraphics[]{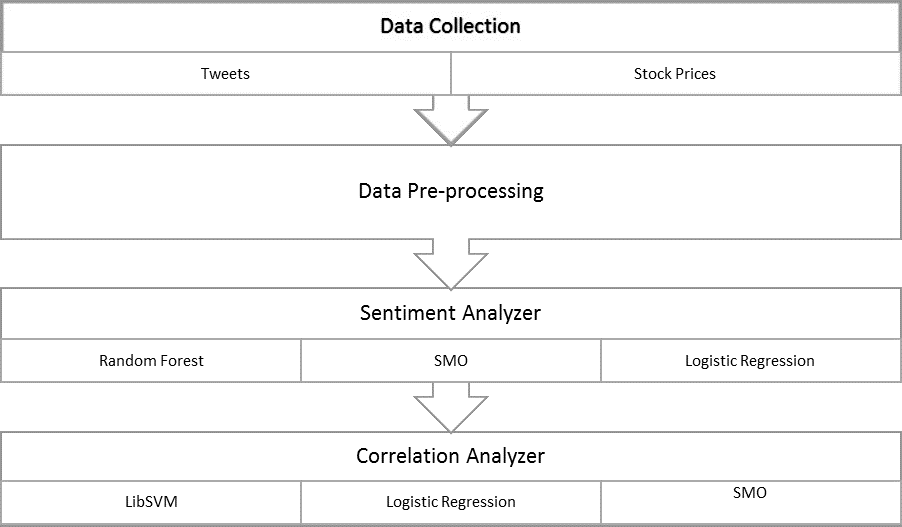}
  \caption{Flow Chart of the proposed analysis}
  \label{fig: Flow Chart of the proposed analysis}
\end{figure*}

\begin{table*}[h]
\centering
\caption{Sample tweets sentiment labeling by the model}
\label{my-label}
\begin{tabular}{|l|l|}
\hline
I'm really excited that today is my first day at @Microsoft as a Technical Evangelist                                                         & 1 \\ \hline
Ultrabooks in the mainstream ($AAPL iPad Pro in contention) | $ MSFT                                                                          & 1 \\ \hline
About the Surface team, it's one of the most secret team at @Microsoft , except VPs, no one knows what they do, so don't expect leaks ;)      & 1 \\ \hline
We broke all your devices, we force you to update whenever we want, and we spy on you all the time. Happy Anniversary!" - @microsoft          & 0 \\ \hline
My grandfather, a depression era company man, hated that \$ MSFT didn't pay dividends (90s). I never could explain "new" idea of growth/cash. & 2 \\ \hline
Thanks @Microsoft , delivered my Surface Pro 3 power plug to a newsagents 20 miles from my home with no notification.                         & 0 \\ \hline
10 swift lessons. My post on @SwiftKey @Microsoft @georgewhitehead @OctopusVentures @IndexVentures @Accel                                     & 0 \\ \hline
@Microsoft Acquires MinecraftEdu, Tailored for Schools http:// nyti.ms/1U9a1oV (via: @nytimesbusiness                                         & 0 \\ \hline
My surface's touchscreen stopped working... I have homework... @Microsoft expand your customer service hours please                           & 0 \\ \hline
MSFT trailing revenues are declining since 2015, net income is up thanks to higher margins                                                    & 2 \\ \hline
\end{tabular}
\end{table*}

\begin{figure*}
\centering
\begin{subfigure}{.5\textwidth}
  \centering
  \includegraphics[width=.8\linewidth]{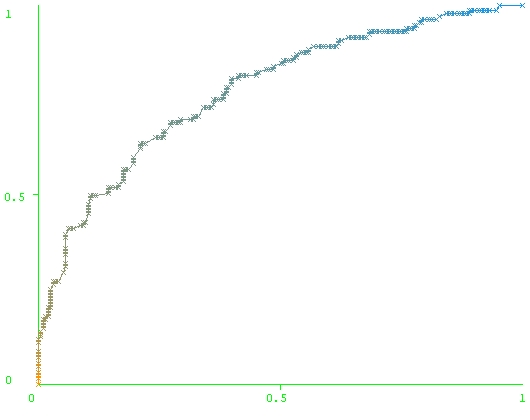}
  \caption{Figure-1\\ROC for Positive sentiment classification\\Area=0.772}
  \label{fig:sub1}
\end{subfigure}%
\begin{subfigure}{.5\textwidth}
  \centering
  \includegraphics[width=.8\linewidth]{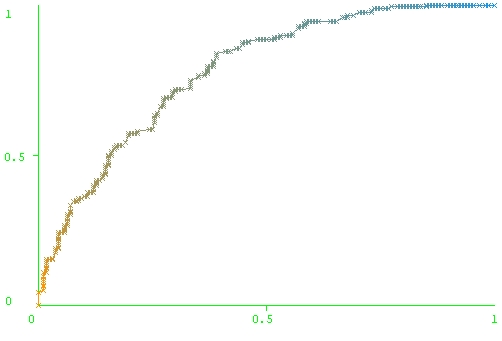}
  \caption{Figure-2\\ROC for Neutral sentiment classification\\Area=0.778}
  \label{fig:sub2}
\end{subfigure}

\begin{subfigure}{.5\textwidth}
  \centering
  \includegraphics[width=.8\linewidth]{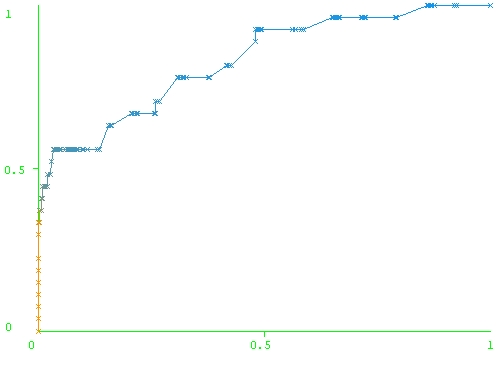}
  \caption{Figure-3\\ROC for Negative sentiment classification\\Area=0.828}
  \label{fig:sub1}
  \end{subfigure}
\end{figure*}

\section{Sentiment Analysis}

Sentiment analysis task is very much field specific. There is lot of research on sentiment analysis of movie reviews and news articles and many sentiment analyzers are available as an open source. The main problem with these analyzers is that they are trained with a different corpus. For instance, Movie corpus and stock corpus are not equivalent. So, we developed our own sentiment analyzer.\\Tweets are classified as positive, negative and neutral based on the sentiment present~\cite{Agarwal}. 3,216 tweets out of the total tweets are examined by humans and annotated as 1 for Positive, 0 for Neutral and 2 for Negative emotions. For classification of nonhuman annotated tweets a machine learning model is trained whose features are extracted from the human annotated tweets.

\subsection{Feature Extraction}
Textual representations are done using two methods:n-grams and Word2vec
\subsubsection{N-gram Representation}
N-gram representation is known for its specificity to match the corpus of text being studied. In these techniques a full corpus of related text is parsed which are tweets in the present work, and every appearing word sequence of length n is extracted from the tweets to form a dictionary of words and phrases. For example the text ``Microsoft is launching a new product" has the following 3-gram word features:``Microsoft is launching", ``is launching a", ``launching a new" and ``a new product". In our case, N-grams for all the tweets form the corpus. In this representation, tweet is split into N-grams and the features to the model are a string of 1’s and 0’s where 1 represents the presence of that N-gram of the tweet in the corpus and a 0 indicates the absence.

\subsubsection{Word2vec Representation}
Word2vec representation is far better, advanced and a recent technique which functions by mapping words to a 300 dimensional vector representations. Once every word of the language has been mapped to a unique vector, vectors of words can be summed up yielding a resultant vector for any given collection of words~\cite{Tomas}.  Relationship between the words is exactly retained in this form of representation. Word vectors difference between Rome and Italy is very close to the difference between vectors of France and Paris This sustained relationship between word concepts makes word2vec model very attractive for textual analysis. In this representation, resultant vector which is sum of 300 dimensional vectors of all words in a tweet acts as features to the model.

\subsection{Model Training}
The features extracted using the above methods for the human annotated tweets are fed to the classifier and trained using random forest algorithm. Both the textual representations performed well and the results are comparable. Out of the two, model trained with word2vec representation is picked because of its sustainability of meaning and promising performance over large datasets. The results of sentiment classification are discussed in the following sections.
The devised classifier is used to predict the emotions of non-human annotated tweets. Table-1 shows a sample of annotated tweets by the sentiment analyzer.

\begin{table*}[h]
\centering
\caption{Sentiment Analysis Results}
\label{my-label}
\begin{tabular}{|l|l|l|l|l|l|l|l|l|}
\hline
\multirow{2}{*}{\begin{tabular}[c]{@{}l@{}}Machine Learning\\ Algorithm\end{tabular}} & \multicolumn{4}{c|}{Word2vec}             & \multicolumn{4}{c|}{N-gram}               \\ \cline{2-9} 
                                                                                      & Accuracy & Precision & Recall & F-Measure & Accuracy & Precision & Recall & F-Measure \\ \hline
Random Forest                                                                         & 70.18\%  & 0.711     & 0.702  & 0.690     & 70.49\%  & 0.719     & 0.705  & 0.694     \\ \hline
Logistic Regression                                                                   & 62.42\%  & 0.621     & 0.624  & 0.621     & 57.14\%  & 0.580     & 0.571  & 0.574     \\ \hline
SMO                                                                                   & 62.42\%  & 0.617     & 0.624  & 0.618     & 65.84\%  & 0.658     & 0.658  & 0.657     \\ \hline
\end{tabular}
\end{table*}

\section{Correlation Analysis of Price and Sentiment }
The stock price data of Microsoft are labeled suitably for training using a simple program. If the previous day stock price is more than the current day stock price, the current day is marked with a numeric value of 0, else marked with a numeric value of 1. Now, this correlation analysis turns out to be a classification problem. The total positive, negative and neutral emotions in tweets in a 3 day period are calculated successively which are used as features for the classifier model and the output is the labeled next day value of stock 0 or 1.The window size is experimented and best results are achieved when the sentiment values precede 3 days to the stock price. A total of 355 instances, each with 3 attributes are fed to the classifier with a split proportions of 80\% train dataset and the remaining dataset for testing. The accuracy of the classifier is discussed in the results section.

\section{Results and Discussion}
This section gives an overview of accuracy rates of the trained classifiers. All the calculations are done in Weka tool which runs on java virtual machine~\cite{Weka}
\subsection{Sentiment Analyzer Results}

The above sections discussed the method followed to train the classifier used for sentiment analysis of tweets. The classifier with features as Word2vec representations of human annotated tweets trained on Random Forest algorithm with a split percentage of 90 for training the model and remaining for testing the model showed an accuracy of 70.2\%. With N-gram representations, the classifier model with same algorithm and with same dataset showed an accuracy of 70.5\%. Though the results are very close, model trained with word2vec representations is picked to classify the nonhuman annotated tweets because of its promising accuracy for large datasets and the sustainability in word meaning. Numerous studies have been conducted on people and they concluded that the rate of human concordance, that is the degree of agreement among humans on the sentiment of a text, is between 70\% and 79\%~\cite{Sentdex}. They have also synthesized that sentiment analyzers above 70\% are very accurate in most of the cases. Provided this information, the results we obtained from the sentiment classification can be observed as very good figures while predicting the sentiments in short texts, tweets, less than 140 characters in length. Table-2 depicts the results of sentiment classification including accuracy, precision, F-measure and recall when trained with different machine learning algorithms. ROC curves are plotted for detailed analysis.

\subsection{Stock Price and Sentiment Correlation Results}
A classifier is presented in the previous sections that is trained with aggregate sentiment values for 3-day period as features and the increase/decrease in stock price represented by 1/0 as the output. Total data is split into two parts, 80 percent to train the model and remaining for testing operations. The classifier results show an accuracy value of 69.01\% when trained using Logistic regression algorithm and the accuracy rate varied with the training set. When the model with LibSVM is trained with 90 percent of data, it gave a result of 71.82\%. These results give a significant edge to the investors and they show good correlation between stock market movements and the sentiments of public expressed in twitter. This trend shows that with increasing dataset the models are performing well. We would like to incorporate more data in our future work.

\section{Conclusion}
In this paper, we have shown that a strong correlation exists between rise/fall in stock prices of a company to the public opinions or emotions about that company expressed on twitter through tweets. The main contribution of our work is the development of a sentiment analyzer that can judge the type of sentiment present in the tweet. The tweets are classified into three categories: positive, negative and neutral. At the beginning, we claimed that positive emotions or sentiment of public in twitter about a company would reflect in its stock price. Our speculation is well supported by the results achieved and seems to have a promising future in research.

\section{Future Work}
In this work, we have considered only twitter data for analyzing people's sentiment which may be biased because not all the people who trade in stocks share their opinions on twitter. Stocktwits~\cite{Stocktwits} is a financial communication platform designed solely for sharing ideas and insights of investors, entrepreneurs and traders. The current study can be extended by incorporating stocktwits data. In addition to this, data from news can also be included for an exhaustive public opinion collection.\\ While training the sentiment analyzer, 3,216 tweets are used which is comparatively a less number to train a sentiment analyzer. In future, we look forward to human annotate more than 10,000 tweets and train the classifiers. With increasing size of training datasets, the models tend to perform better.


\section*{Acknowledgment}

The authors would like to thank the students of IIT Bhubaneswar who contributed to the human annotation of tweets.



%

\end{document}